\documentclass[twocolumn,prb,amsmath,amssymb,floatfix,superscriptaddress,showpacs]{revtex4-1}
\usepackage{graphicx}
\usepackage{dcolumn}
\usepackage{bm}
\usepackage{color}
\usepackage{xspace}
\providecommand*{\tc}{$T_c$\xspace}%

\def\ybco{YBa$_2$Cu$_3$O$_{6+x}$}
\def\lesco{La$_{1.8-x}$Eu$_{0.2}$Sr$_x$CuO$_4$}
\def\lbco{La$_{2-x}$Ba$_x$CuO$_4$}
\def\lsco{La$_{2-x}$Sr$_x$CuO$_4$}
\def\bscco{Bi$_2$Sr$_2$CaCu$_2$O$_{8+\delta}$}

\begin{document}

\title{Probing the connections between superconductivity, stripe order, and structure in La$_{1.905}$Ba$_{0.095}$Cu$_{1-y}$Zn$_y$O$_4$}
\author{Jinsheng Wen}
\affiliation{Condensed Matter Physics and Materials Science
Department, Brookhaven National Laboratory, Upton, New York 11973,
USA}
\affiliation{Department of Physics,
University of California, Berkeley, California 94720, USA.}
\author{Zhijun Xu}
\affiliation{Condensed Matter Physics and Materials Science
Department, Brookhaven National Laboratory, Upton, New York 11973,
USA} \affiliation{Physics Department, The City College of New York,
New York, New York 10031, USA}
\author{Guangyong Xu}
\affiliation{Condensed Matter Physics and Materials Science
Department, Brookhaven National Laboratory, Upton, New York 11973,
USA}
\author{Qing Jie}
\affiliation{Condensed Matter Physics and Materials Science
Department, Brookhaven National Laboratory, Upton, New York 11973,
USA} \affiliation{Department of Materials Science and Engineering,
Stony Brook University, Stony Brook, New York 11794, USA}
\author{M. H\"ucker}
\affiliation{Condensed Matter Physics and Materials Science
Department, Brookhaven National Laboratory, Upton, New York 11973,
USA}
\author{A.~Zheludev}
\altaffiliation{Present address: Neutron Scattering and Magnetism,
Institute for Solid State Physics, ETH Zurich, Switzerland.}
\affiliation{Neutron Scattering Science Division, Oak Ridge National Laboratory, Oak Ridge, Tennessee 37831, USA}
\author{Wei~Tian}
\affiliation{Ames Laboratory and Department of Physics and Astronomy, Iowa State University, Ames, Iowa 50011, USA}
\author{B.~L.~Winn}
\altaffiliation{Present address: Neutron Scattering Science Division, Oak Ridge National Laboratory, Oak Ridge, Tennessee 37831, USA.}
\affiliation{Condensed Matter Physics and Materials Science Department, Brookhaven National Laboratory, Upton, New York 11973, USA}
\author{J.~L.~Zarestky}
\affiliation{Ames Laboratory and Department of Physics and Astronomy, Iowa State University, Ames, Iowa 50011, USA}
\author{D.~K.~Singh}
\affiliation{NIST Center for Neutron Research, National Institute of
Standards and Technology, Gaithersburg, Maryland 20899, USA}
\affiliation{Department of Materials Science and Engineering, University of Maryland, College Park, Maryland 20742, USA}
\author{Tao Hong}
\affiliation{Quantum Condensed Matter Division, Oak Ridge National Laboratory, Oak Ridge, Tennessee 37831, USA}
\author{Qiang Li}
\affiliation{Condensed Matter Physics and Materials Science
Department, Brookhaven National Laboratory, Upton, New York 11973,
USA}
\author{Genda Gu}
\affiliation{Condensed Matter Physics and Materials Science
Department, Brookhaven National Laboratory, Upton, New York 11973,
USA}
\author{J.~M.~Tranquada}
\affiliation{Condensed Matter Physics and Materials Science Department, Brookhaven National Laboratory, Upton, New York 11973, USA}
\date{\today}

\begin{abstract}
The superconducting system \lbco\ is known to show a minimum in the transition temperature, $T_c$, at $x=\frac18$ where maximal stripe order is pinned by the anisotropy within the CuO$_2$ planes that occurs in the low-temperature-tetragonal (LTT) crystal structure.  For $x=0.095$, where $T_c$ reaches its maximum value of 32~K, there is a roughly coincident structural transition to a phase that is very close to LTT.  Here we present a neutron scattering study of the structural transition, and demonstrate how features of it correlate with anomalies in the magnetic susceptibility, electrical resistivity, thermal conductivity, and thermoelectric power.  We also present measurements on a crystal with 1\%\ Zn substituted for Cu, which reduces $T_c$ to 17~K, enhances the spin stripe order, but has much less effect on the structural transition.  We make the case that the structural transition correlates with a reduction of the Josephson coupling between the CuO$_2$ layers, which interrupts the growth of the superconducting order. We also discuss evidence for two-dimensional superconducting fluctuations in the normal state, analyze the effective magnetic moment per Zn impurity, and consider the significance of the anomalous thermopower often reported in the stripe-ordered phase.
\end{abstract}


\maketitle

\section{Introduction}

The existence of charge and spin stripe order in some underdoped cuprate superconductors is now well established.\cite{zaan01,kive03,lee06,vojt09}  The spin and charge orders have been detected directly by neutron and x-ray diffraction,\cite{tran95a,vonz98,ichi00,fuji02,fuji04,abba05,huck07,fink09,wilk11} charge-stripe-like density-of-states modulations have been detected by scanning tunneling spectroscopic imaging,\cite{howa03b,hoff02,kohs07,park10} and related nematic responses have been reported.\cite{waki00,mook00,stoc04,hink08,haug10,daou10,lawl10}  The controversial issue is whether stripe order is simply a phase that competes with superconductivity,\cite{rice97,scal01,lee08} or whether it might be symptomatic of the pairing mechanism.\cite{kive03,emer97,hime02,deml04,berg09b,whit09,lode11}  Stripe order is strongest at $x=1/8$ in La$_{2-x}$Ba$_x$CuO$_4$~(LBCO) and closely related cuprates,\cite{ichi00,huck11,fink11} where it also corresponds to a local minimum\cite{mood88,koik91b,koik92,craw91} in $T_c$ for bulk superconductivity.

Despite the competition with bulk superconductivity, recent work\cite{li07,tran08} on LBCO with $x=1/8$ has provided evidence for the development of strong two-dimensional (2D) superconducting correlations for $T\lesssim 40$~K.  It appears that stripe order does not compete with pairing correlations within the CuO$_2$ planes, but instead frustrates the Josephson coupling between the layers,\cite{taji01} which inhibits the development of 3D superconducting order.  To explain this effect, it has been proposed that pair-density-wave superconductivity develops, with strong pairing in each charge stripe and an oscillation of the phase of the superconducting order parameter from one stripe to the next.\cite{hime02,berg09b,berg07,berg09a}

\begin{figure}[t]
\includegraphics[width=0.8\linewidth]{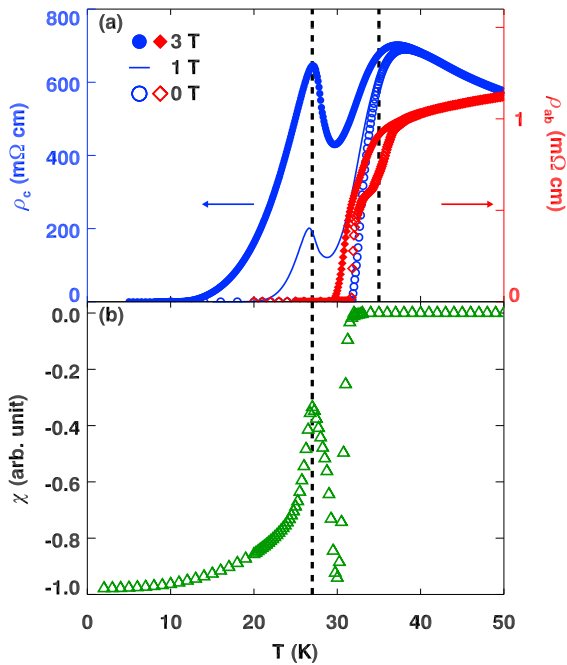}
\caption{(Color online)  (a) Plots of $\rho_{ab}$ (diamonds) and $\rho_c$ (circles, line) for LBCO $x=0.095$ in $\mu_0H_\bot=0$~T (open symbols), 1~T (line), and 3~T (filled symbols), from paper I.  (b)  Field-cooled susceptibility measured in $\mu_0H_\bot=0.01$~T, from Ref.~\onlinecite{huck11}.  The vertical dashed lines mark 27~K and 35~K.}
\label{fig:spike}
\end{figure}

As the stripe order in LBCO has its maximum amplitude at $x=1/8$,\cite{huck11} that composition has received the greatest attention in terms of scattering studies\cite{fuji02,fuji04,abba05,kim08,kim08b,wen08b,huck10,wilk11}; however, stripe order has also been observed in samples with $\Delta x=\pm0.03$.\cite{huck11,fuji05,duns08}  The reduction in stripe order correlates with a rise in the bulk $T_c$,  with $T_c$ on the underdoped side reaching a maximum of 32~K for $x=0.095$. In the previous paper,\cite{wen11} which we will refer to as paper I, we have addressed the impact of an applied magnetic field on resistivity and stripe order in LBCO with $x=0.095$. Intriguing effects are observed when a magnetic field $H_\bot$  is applied perpendicular to the planes, as illustrated in Fig.~\ref{fig:spike}. The most significant feature is that the field induces finite resistivity perpendicular to the planes ($\rho_c$), but has little impact on the drop in in-plane resistivity ($\rho_{ab}$), at least for fields up to 9~T. Analysis in paper I indicates that  thermally-induced phase fluctuations cause $\rho_c$ to be finite even when the interlayer Josephson coupling is still finite.  (That analysis provides new insights regarding the behavior of the $x=1/8$ sample, as we will discuss.)  The most striking feature in Fig.~\ref{fig:spike} is the spike in $\rho_c$ at 27~K when $H_\bot$ is finite; a similar effect is also seen in the bulk susceptibility.  This behavior looks very similar to that found in GdMo$_6$S$_8$, where the effect was attributed to pair-breaking effects associated with the antiferromagnetic ordering of the Gd moments.\cite{ishi82}  The physics appears to be different in the present case, as we do not have any significant magnetic ordering.  Instead, we will show that there is a structural transition that coincides with the superconducting transition.  The low-temperature phase is similar to that in the LBCO $x=1/8$ composition, and we will argue that the Josephson coupling is reduced by the transition, leading to the anomalous behavior of $\rho_c$.  The temperature-dependent growth of the superfluid density is also sensitive to the Josephson coupling, and this leads to the unusual features in $\chi(T)$.  The structural transition actually begins at $\sim35$~K, corresponding to the plateau in $\rho_{ab}$ seen in zero field.

To provide a justification for this interpretation, we present detailed neutron diffraction characterizations of the phase transition to the low-temperature structure, which pins (weakly, for $x=0.095$) charge stripe order.  In addition, we present measurements of the in-plane thermal conductivity and thermoelectric power.  As a further test, we compare results for a second crystal in which 1\%\ Zn has been substituted for Cu; the Zn depresses the superconducting $T_c$, but enhances the stripe order without changing its onset temperature.  The impact of the Zn on the spin-stripe order is similar to the application of $H_\bot$; however, its impact on the superconductivity is much more deleterious.

Looking at the spin susceptibility above $T_c$, we find evidence for 2D superconducting fluctuations.  Zn impurities do not destroy these fluctuations, but they do shift the divergence of the fluctuations to a reduced temperature.  We also analyze the Curie-like contribution to the susceptibility by effective magnetic moments induced by the Zn impurities.

The impact of field and Zn-doping on the thermopower, $S$, is of particular interest. Taillefer and coworkers\cite{chan10,lali11} have recently pointed out strong similarities in $S(T)$ between underdoped \ybco\ (YBCO) in strong magnetic field (to suppress superconductivity) and stripe-ordered compounds, such as \lesco\ and LBCO.  While the phenomenological connection is quite appealing, they make a bold claim that the temperature dependence of the thermopower in all of these compounds is a consequence of Fermi-surface reconstruction.  We provide an alternative analysis.  In particular, we show for LBCO $x=0.095$ that the drop in $S/T$ on cooling is associated with pairing correlations.  We also discuss the case of LBCO $x=1/8$.

The rest of the paper is organized as follows.  The experimental methods are described in the following section.  The results are presented in Sec.~III, where we cover the structural transitions, spin-stripe order, superconducting transitions, and thermal conductivity and thermopower.  Section IV contains a discussion of these results and their implications.  Conclusions are presented in Sec.~V.

\section{Experimental Methods}

The crystals were grown in an infrared image furnace by the floating-zone technique. Single crystals of La$_{1.905}$Ba$_{0.095}$CuO$_4$ (Zn0) and La$_{1.905}$Ba$_{0.095}$Cu$_{0.99}$Zn$_{0.01}$O$_4$ (Zn1) used for the neutron scattering studies were identified and extracted with the help of polarized-light microscopy.  The 11.2-g Zn0 crystal has been studied with neutrons in paper I and earlier\cite{huck11}; crystals for transport and susceptibility measurements  were extracted from the same as-grown crystal rod.  The Zn1 crystal for the neutron experiment has a mass of 4.1~g, with secondary crystals cut from the same rod.  X-ray diffraction measurements on a secondary crystal were reported in Ref.~\onlinecite{huck11b}.

Data for thermal conductivity and thermoelectric power were collected at the same time in a Physical Property Measurement System (PPMS) with the thermal transport option by the four-probe dc steady state method. The temperature gradient along the $a$-$b$ plane is 1\%\ of the temperature across the crystal. The Zn0 sample is 1.3~mm high $\times$ 0.8~mm thick, with 2.9~mm between the contacts measuring the voltage and temperature gradient. This sample was also used in paper I to measure the $a$-$b$-plane resistivity. The Zn1 piece is 1.0~mm high $\times$ 0.3~mm thick, with 3.5~mm between voltage contacts.   The bulk susceptibility $\chi$ was measured with a magnetometer equipped with a superconducting-quantum-interference-device (SQUID).  Measurements of the anisotropic spin susceptibility of both samples were reported in Ref.~\onlinecite{huck11b}.

The neutron scattering experiments were performed on the triple-axis spectrometers HB1 and HB1A, both located at the High Flux Isotope Reactor, Oak Ridge National Laboratory, and SPINS, located at the NIST Center for Neutron Research, National Institute of Standards and Technology. Beam collimations used on the three spectrometers are $48'$--$60'$--$S$--$60'$--$240'$, $48'$--$48'$--$S$--$40'$--$136'$, and $55'$--$80'$--$S$--$80'$--$240'$ ($S=$ sample), respectively. The measurements were performed in a fixed final energy ($E_f$) mode with $E_f=$14.7~meV on HB1 and HB1A, and $E_f=5$~meV on SPINS.  To reduce contamination from higher-order neutrons, pyrolytic graphite filters were put before and after the sample on HB1 and HB1A, and a cooled Be filter was put after the sample on SPINS for elastic measurements. The inelastic measurements were performed on HB1. The samples were mounted such that the $(HK0)$ zone was parallel to the scattering plane, with $(HK0)$ defined by vectors [100] and [010] in orthorhombic notation [see Fig.~\ref{fig:twin}(c)]. Neutron scattering data are described in terms of reciprocal lattice units (rlu) of $(a^*,b^*,c^*)=(2\pi/a,2\pi/b,2\pi/c)$, where $a=b=5.35$~\AA\ and $c=13.2$~\AA\ at base temperature. In the magnetic-field measurements on SPINS, a vertical-field superconducting magnet was used, so that the field was applied perpendicular to the $a$-$b$ plane. The intensity of the spin stripe order has been normalized to the results of the LBCO $x = 1/8$ sample
in zero field at low temperature using measurements of an acoustic phonon mode near the (200) Bragg peak. Cross normalizations using incoherent elastic scattering and sample mass have also been done and give consistent results.

\section{Results}

\subsection{Structural transitions}

\begin{figure}[b]
\includegraphics[width=0.8\linewidth]{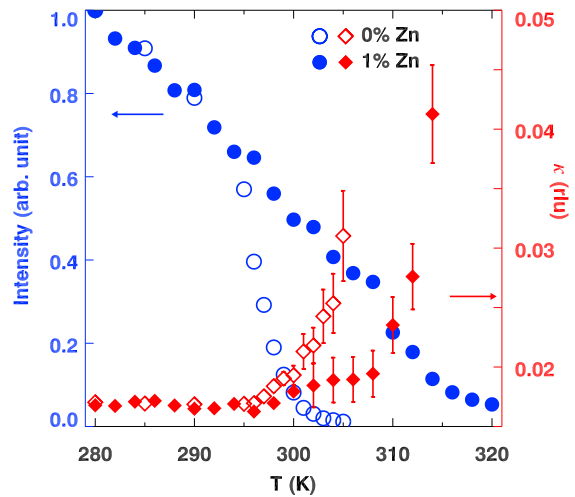}
\caption{(Color online) Peak intensity (left axis, circles) and half width at half maximum (right axis, diamonds) as a function of temperature for the (032) superlattice peak at temperatures near 300~K for the Zn0 (open symbols) and Zn1 (filled symbols) samples. Error bars represent $\pm$1 standard deviation throughout the paper.}
\label{fig:032}
\end{figure}

Each of our samples undergoes two structural transitions on cooling from above room temperature, as described in Ref.~\onlinecite{huck11}.  The transition from the high-temperature-tetragonal (HTT, $I4/mmm$ symmetry) phase to low-temperature-orthorhombic (LTO, $Bmab$) occurs very close to room temperature for $x\sim0.095$ and is sensitive to the Ba concentration.  (The transition temperature shifts by $\sim23$~K per 0.01~Ba content.\cite{huck11,axe89})  We consider it here in order to evaluate the impact of Zn impurities in the Zn1 sample.  To determine the LTO-to-HTT transition, we measured the (032) superlattice peak of the LTO phase on warming.  The peak intensity and {\bf Q} width are plotted in Fig.~\ref{fig:032} for both samples.  The second order transition is rounded due to disorder.  From these data, we estimate the transition temperatures to be $298\pm3$~K and $308\pm5$~K for Zn0 and Zn1, respectively.   If we associate the difference in  transition temperatures with Ba concentration, it suggests a smaller Ba concentration by 0.004(2) for the Zn1 sample.\cite{huck11b}

\begin{figure}[t]
\includegraphics[width=0.8\linewidth]{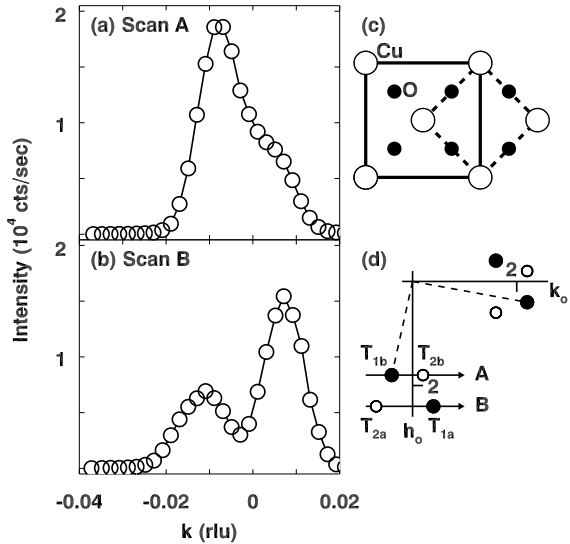}
\caption{(a) and (b) Two scans through the LTO peaks of the Zn0 crystal at 54~K with trajectories indicated in (d). (c) Schematic of the orthorhombic (solid line), and high-temperature tetragonal (dashed line) unit cells; large and small circles indicate Cu and O site respectively. In this work, the orthorhombic notation is used. In (d), the filled and open circles indicate the strong and weak, respectively, (200) and (020) Bragg peaks that are present simultaneously due to twinning.  The strong twin pair is labelled T$_{1a}$ and T$_{1b}$; the weak pair are T$_{2a}$ and T$_{2b}$.  Arrows indicate directions of scans in (a) and (b); the dashed lines connect a pair of (200) and (020) peaks belonging to a single domain. }
\label{fig:twin}
\end{figure}

In transforming from HTT to LTO, an LBCO crystal develops twin domains.\cite{zhu94}  A typical twin domain  boundary is oriented along a line of Cu-O nearest-neighbor bonds, with the orthorhombic $a$ (or $b$) axis of one domain reflected symmetrically about the boundary into the twin domain.  Since the Cu-O bond direction is along a diagonal of the orthorhombic cell, the twin boundary results in an angle between the $a$ axes of neighboring twin domains that deviates slightly from $90^\circ$.   In terms of the orthorhombic strain, $s=2(b-a)/(b+a)$, the deviation from a right angle is approximately $s$ radians.  For diffraction, this means that the (200) reflection of twin domain T$_{1a}$ will occur in almost the same direction as the (020) of twin domain T$_{1b}$, but with an angular separation of $s$ radians [see Fig.~\ref{fig:twin}(d)].

In general, there will be more than one pair of twin domains.  (The twin pattern depends on strain effects, which can depend on crystal size.)  As there are two Cu-O bond directions that can orient twin boundaries, reflections from two pairs of twin domains (for a total of four) are typically observed.  For our Zn0 sample, we observe two such pairs, but with an unusual angle between the pairs.  Scans through a set of (200)/(020) reflections are shown in Fig.~\ref{fig:twin}(a) and (b), with schematic illustration of the peak and scan orientations in (d).  The twin pair T$_1$ with greater intensity corresponds to peaks at $(1.998,-0.007,0)$ and $(2.014,0.009,0)$, with the weaker pair, T$_2$, appearing at $(1.998,0.004,0)$ and $(2.014,-0.012,0)$.  For each pair, $s=0.008$, consistent with our previous x-ray diffraction results.\cite{huck11}

\begin{figure}[t]
\includegraphics[width=0.8\linewidth]{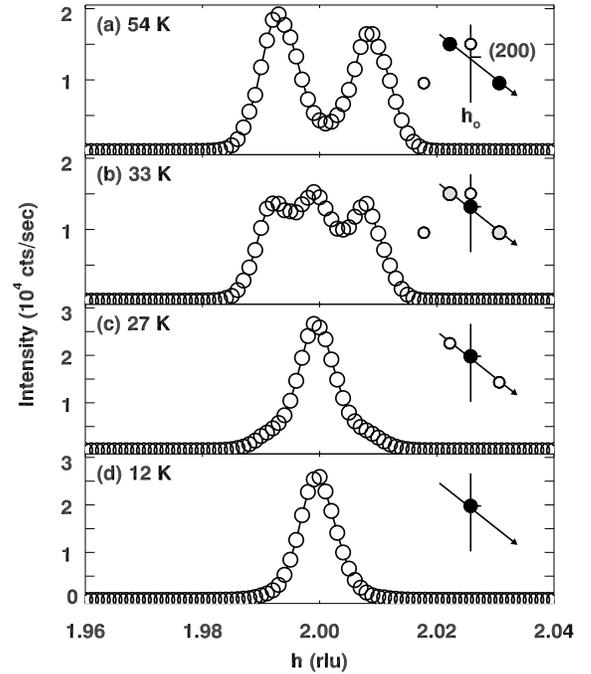}
\caption{Diagonal scans through the T$_1$ peaks of the LTO phase of the Zn0 crystal at temperatures of (a) 54~K, (b) 33 K, (c) 27 K, and (d) 12), showing the evolution from the split peaks of the LTO phase to the single peak of the LTLO phase. Insets show the schematics of the peak patterns, and the scan direction; closed circles: strong peaks; grey circles: weaker peaks; open circles, weakest peaks.}
\label{fig:Tscan}
\end{figure}

The reason for the fuss here about twin domains is that they are relevant to measuring the transition from the LTO phase to the low-temperature-less-orthorhombic (LTLO, $Pccn$). (Note that we adopt this notation from the X-ray diffraction measurements,\cite{huck11} which have shown that the low temperature phase for $x=0.095$ is slightly orthorhombic, corresponding to space group $Pccn$; however, the residual orthorhombic strain, which is only 3\%\ of that in the LTO phase, could not be resolved in the neutron measurements.) The (200) reflection of the tetragonal phase appears close to the half-way point between an orthorhombic (200)/(020) twin pair.  We illustrate this in Fig.~\ref{fig:Tscan} with a scan through the T$_1$ reflections [see Fig.~\ref{fig:twin}(d) and the inset of Fig.~\ref{fig:Tscan}], repeated at several temperatures.  At 54~K, we see only the two peaks of the LTO phase, while at 33~K a significant LTLO peak is present in the middle.  By 27~K, the LTLO peak has grown stronger, with the remnant LTO peaks appearing as small shoulders.  Finally, the LTO peaks are absent at 12~K.

\begin{figure}[t]
\includegraphics[width=0.8\linewidth]{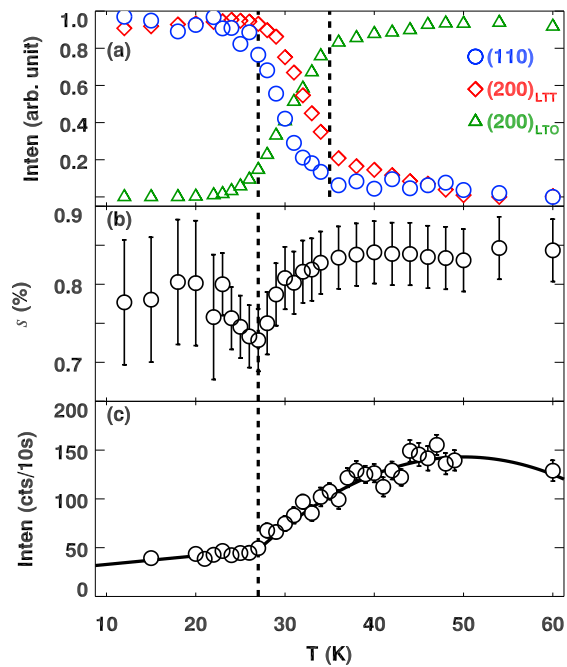}
\caption{(Color online) (a) Normalized intensities of the (200) peaks [LTO T$_{1b}$ peak (triangles), LTLO  peak (diamonds)] and the (110) superlattice peak of the LTLO phase (circles) as a function of temperature.  Vertical dashed lines indicate the main region of the transition. (b) Orthorhombic strain vs.\ temperature.   (c) Temperature dependence of the inelastic scattering intensity measured at ${\bf Q}= (0.01,3,2)$ and $\hbar\omega= 1$~meV.  The solid line is a guide to the eye.}
\label{fig:td2}
\end{figure}

To quantify the transition, we have fit a set of 3 gaussian peaks to scans of the type shown in Fig.~\ref{fig:Tscan}.  In Fig.~\ref{fig:td2}(a), we plot the temperature dependent intensities of representative (200) peaks and the (110) superlattice peak which is finite only in the LTLO phase.  As one can see, the two phases coexist over a significant temperature range.  The $(200)_{\rm LTO}$ peaks appear to start transferring intensity to the $(200)_{\rm LTLO}$ below 50~K, but the main transition does not begin until about 35~K.  Interestingly, the (110) superlattice peak intensity comes up more slowly.  The temperature at which the transition is completed is not so clear from the peak intensities.  To provide another measure, Fig.~\ref{fig:td2}(b) shows the temperature dependence of the orthorhombic strain obtained from the (200) peak fits.  It decreases below 35~K, and shows a sharp minimum at 27~K, where most of the LTO phase has transformed.

We have also attempted to measure critical scattering associated with the transition.  To do this, we positioned the spectrometer at ${\bf Q}=(0.01,3,2)$ with energy transfer $\hbar\omega=1$~meV and measured the scattering rate as the temperature was varied; the results are shown in Fig.~\ref{fig:td2}(c).  There might be a broad peak near 50~K, where the transition begins.  In any case, the scattering from a soft phonon decreases with temperature, with the intensity showing a kink at 27~K.  We will see shortly that this corresponds well with temperature dependence of the thermal conductivity.

\begin{figure}[t]
\includegraphics[width=0.85\linewidth,trim=-7mm 0mm -1mm 118mm,clip]{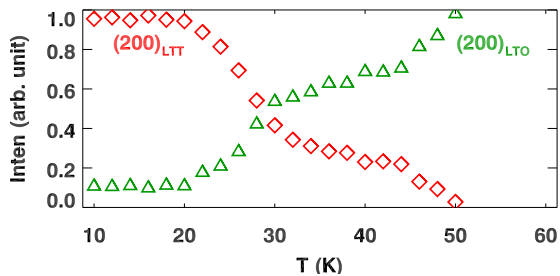}
\caption{(Color online) LTO to LTLO transition in the Zn1 sample.  Normalized peak intensities for a $(200)_{\rm LTO}$ peak (triangles) and the $(200)_{\rm LTLO}$ (diamonds) as a function of temperature.}
\label{fig:td2zn}
\end{figure}

Figure~\ref{fig:td2zn} shows the temperature dependence of LTO and LTLO (200) peak intensities for the Zn1 sample.  Here we see that the transition appears to start on cooling through 50~K.  The transformation is gradual at first, and then more rapid below 30~K.  We will see later that thermal conductivity measurements indicate that the completion of the transition is at 27~K, as for the Zn0 sample. In contrast, such a result is not obvious from the diffraction data.

In this part, we have demonstrated that in LBCO ($x=0.095$), an LTO-LTLO transition starts at 35~K, and completes at 27~K. The 1\% Zn doping does not have much impact on the structural transition.

\subsection{Spin-stripe order}

Our previous studies (paper I and Ref.~\onlinecite{huck11}) have shown that in Zn0 weak spin- and charge-stripe order develop below $\sim30$~K.  The integrated intensity for the spin-stripe-order peak, proportional to the square of the spatially-modulated ordered moment times the ordered volume fraction, is only a tenth of that found for LBCO $x=1/8$. The correlation length, estimated from scans performed along the $h$ direction, is $\sim120$~\AA{}, which is about a half of that of the latter sample. Spin and charge orders are both found to be enhanced by a transverse magnetic field.

\begin{figure}[t]
\includegraphics[width=0.9\linewidth]{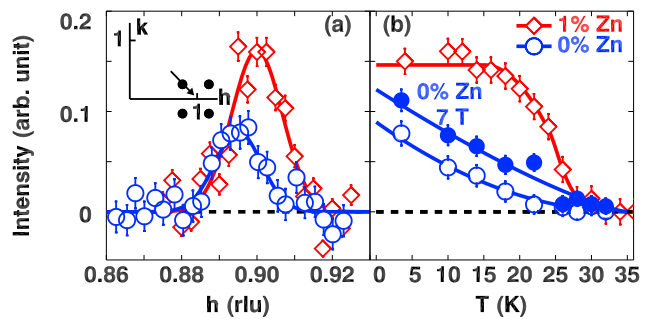}
\caption{(Color online)  (a) Scans through one of the spin-stripe-order peaks along the direction shown in the inset for Zn1 (diamonds) and Zn0 (circles) at $T=5$~K. To remove background, scans performed at 50~K have been subtracted. Lines through data are gaussian peak fits.  (b) Integrated intensity of the spin-stripe-order peak as a function of temperature for the two samples. The intensities for the Zn0 sample in a $c$-axis magnetic field are indicated by filled circles (from paper I).  Error bars reflect counting statistics.  Lines through the data are guides to the eye.}
\label{fig:spin}
\end{figure}

In the present work, we have measured superlattice peaks arising from the spin-stripe order in the Zn1 sample. The magnetic peaks appear at positions $(1\pm\epsilon,\pm\epsilon,0)$ with $\epsilon\approx0.1$, as indicated in the inset of Fig.~\ref{fig:spin}(a).  A scan at 5~K through the $(1-\epsilon,\epsilon,0)$ peak is indicated by the diamonds in Fig.~\ref{fig:spin}(a), with data from the Zn0 sample denoted by circles; the intensities have been normalized to the results from the LBCO $x=1/8$ sample.~\cite{wen08b} The Zn doping has doubled the peak intensity.  This enhancement of the peak intensity is consistent with previous works on Zn doped \lsco\ (LSCO), where it has been observed that Zn impurities increase the spectral weight for low- and zero-energy magnetic scattering.~\cite{tran99b,kimu99,kimu03b,waki05}  The peak width is approximately the same for both samples, indicating that the Zn impurities have not significantly degraded the spin-spin correlation length. On the other hand, the incommensurability is slightly reduced in Zn1 compared to Zn0. This is consistent with the estimated small reduction of Ba concentration in the Zn1 sample and the approximately linear relationship between the incommensurability and $x$ at this hole concentration.~\cite{birg06}

The temperature dependences of the peak intensities are compared in Fig.~\ref{fig:spin}(b). For Zn0, the stripe-order peak intensity increases slowly with cooling, showing a glassy behavior; in contrast, the intensity for Zn1 looks more like the order parameter of a second-order transition.  This curve looks similar to the impact of $H_\bot$ on La$_{1.9}$Sr$_{0.1}$CuO$_4$.\cite{lake02} We might have expected a field to have the same impact on the Zn0 sample, but while the field enhances the intensity, the thermal evolution remains glass-like, as we have shown in paper I. Impact of magnetic fields on the Zn0 sample have been discussed in great details in paper I. Here, for easy comparisons, the field effect on the spin-stripe order of the Zn0 sample is also plotted in Fig.~\ref{fig:spin}(b).

\begin{figure}[t]
\includegraphics[width=0.8\linewidth]{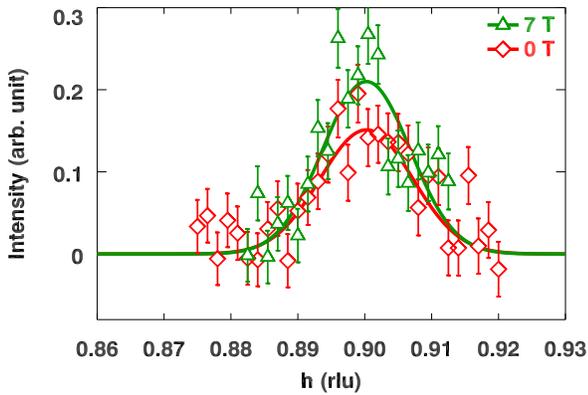}
\medskip
\caption{(Color online)  Scans through $(0.9,0.1,0)$ for the Zn1 sample under zero (diamonds) and 7-T magnetic field (triangles) at 4~K with intensities normalized to the results of the La$_{1.875}$Ba$_{0.125}$CuO$_4$ sample in zero field.~\cite{wen08b}  Background measured at 50~K has been subtracted. Lines through data are fitted gaussians.}
\label{fig:mag}
\end{figure}

We have also applied a $c$-axis magnetic field to the Zn1 sample at $T=4$ K; scans through the magnetic peak with and without a field are compared in Fig.~\ref{fig:mag}. The result in the field was obtained after field cooling from above $T_c$; the alignment was checked by scans through the (200)/(020) Bragg peaks to be sure that the sample orientation was not changed by torque due to the field.\cite{stoc09} Determining peak areas by integrating the measured intensities (after subtracting backgrounds measured at 50~K), we find that the 7-T field increases the integrated intensity by $(37\pm16)$\%, where the error bar is one standard deviation.

There have been many studies on the magnetic-field effect on the spin-stripe order in LSCO.~\cite{lake02,chan08,fuji12} In these samples, spin-stripe order is relatively weak (compared to that of LBCO $x=1/8$) in terms of peak intensity and correlation length. Application of an external field often results in the enhancement of the stripe-order peak intensity, which is interpreted as a result of the competition between stripe order and superconductivity.~\cite{lake02,chan08,fuji12} For LBCO $x=1/8$, where spin-stripe order is maximized, and static charge-stripe order is observed, we have shown that there is still an enhancement by the field, but much less pronounced than those in LSCO~~\cite{lake02,chan08} and LBCO $x=0.095$ samples.~\cite{wen11,wen08b} In Ref.~\onlinecite{duns08} it was reported that an applied field had no effect on the spin-order peak intensity in LBCO $x=0.095$, nominally the same Ba concentration as studied here and in paper I. The apparent conflict is resolved when one compares structural transition temperatures, as differences there indicate that the Ba concentration in the sample studied in Ref.~\onlinecite{duns08} is significantly larger than that of our $x=0.095$ crystals, as discussed in paper I.

The impact of a magnetic field has also been studied in \ybco.  Measurements by Haug {\it et al.}\cite{haug09,haug10} on detwinned crystals with $x=0.35$ and 0.45 revealed an enhancement of the elastic incommensurate magnetic signal; the relative change was larger for $x=0.45$ where the zero-field intensity is weaker.  In contrast, Stock {\it et al.}\cite{stoc09} studied twinned crystals with $x=0.33$ and 0.35, where no field enhancement of elastic magnetic intensity was observed; however, the applied field was found to enhance the inelastic magnetic response for energies less than 1 meV in both samples.


Our observation of a modest increment of the spin-stripe order peak intensity by the magnetic field in Zn1 is consistent with the trend observed in LSCO and LBCO.\cite{lake02,chan08,fuji12} The point we want to make here is that Zn doping enhances the spin-stripe-order peak intensity. The intensity is further enhanced by applying a magnetic field, and the amount of enhancement is comparable to that in the Zn0 sample.

\subsection{Susceptibility}

\begin{figure}[b]
\includegraphics[width=0.8\linewidth]{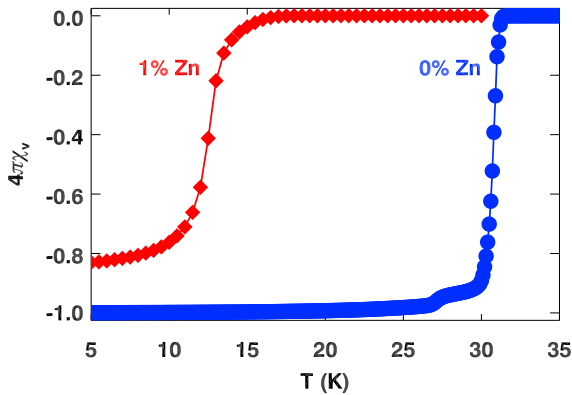}
\medskip
\caption{(Color online) Bulk susceptibility for Zn0 (circles), and Zn1 (diamonds) obtained under zero-field-cooling conditions in a 2-Oe field perpendicular to the $ab$ plane.}
\label{fig:sc}
\end{figure}

To characterize the superconducting transitions of the Zn0 and Zn1 samples, the magnetic susceptibility was measured in a 2-Oe field applied perpendicular to the $ab$ planes after zero-field cooling; the results are shown in Fig.~\ref{fig:sc}. The Zn0 sample has a bulk \tc\ of 32~K, and becomes fully diamagnetic below 26~K.  In the Zn1 sample, \tc is reduced to 17~K, and the screening volume fraction is also reduced. In LSCO with hole concentration close to the optimal doping level, $x=0.155$, \tc is reduced to 16~K by 1.7\%\ Zn,~\cite{kofu09} and to 8~K with 2.5\%\ Zn.~\cite{naka98c}

The suppression of the superconductivity has been successfully explained by the ``Swiss cheese" model,~\cite{nach96} or an identical ``island" model,~\cite{naka98c,hiro01,kofu05} in which each Zn impurity in a CuO$_2$ plane suppresses superconductivity within an area of $\pi\xi^2$, with $\xi$ being the superconducting coherence length, $\xi\sim20$~\AA.~\cite{naka98c} In this sense, the Zn-doping and magnetic-field effects are similar.  As a magnetic field perpendicular to the planes penetrates into the superconductor, it induces non-superconducting vortex cores of area $\pi\xi^2$ within the planes.~\cite{blat94}  For underdoped LSCO, each type of normal core appears to induce a much larger patch of spin stripe order.\cite{birg06}  For our Zn0 sample, we have demonstrated in paper I that the vortices also enhance the charge stripe order.

The susceptibility shown in Fig.~\ref{fig:sc} for Zn0 has a small step at 27~K.  This is a considerably reduced version of the sharp kink shown in Fig.~\ref{fig:spike}(b) for the sample studied in Ref.~\onlinecite{huck11}.  That sample was taken from a different as-grown crystal rod than the present one; nevertheless, the kink and spike are at the same temperature, which appears to correspond to the completion of the LTO to LTLO transition.  We will return to this issue in the discussion section.

We can learn about superconducting fluctuations above the transition by studying the anisotropy of the spin susceptibility.\cite{huck08}  The spin susceptibility, $\chi_i^s(T)$, is sensitive to the orientation of the magnetic field, $i=c$ or $ab$.  It can be obtained from the bulk susceptibility, $\chi_i$, using
\begin{equation}
  \chi_i^s(T) = \chi_i(T) - \chi^{\rm core} - \chi_i^{\rm VV},
\end{equation}
where $\chi^{\rm core}$ is the core diamagnetism\cite{land86} and $\chi_i^{\rm VV}$ is the Van Vleck susceptibility due to spin-orbit coupling between Cu $3d$ states.   As discussed in Ref.~\onlinecite{huck08}, the spin susceptibility is dominated by the paramagnetic response of coupled local moments, with effective moments that depend on anisotropic gyroscopic factors $g_i$.  The quantities $\chi_i^{\rm VV}$ and $g_i$ can be determined by a fit to the data with the assumption that $\chi_i^2/g_i^2$ is isotropic in the paramagnetic regime.  New susceptibility measurements\cite{huck11b} were performed for the Zn0 and Zn1 sample, extending to 375~K.  The Van Vleck susceptibilities were assumed to be the same as in the previous analysis ($\chi_{ab}^{\rm VV}=0.111\times 10^{-4}$ and $\chi_c^{\rm VV}=0.333\times 10^{-4}$ emu/mol), but a new fit for the $g$-factors was performed over the temperature range 150--375~K, yielding $g_{ab}=2.189$ and $g_{c}=2.566$, close to the previous values.  We assume that the same parameters apply to the Zn1 sample.

\begin{figure}[t]
\includegraphics[width=0.8\linewidth]{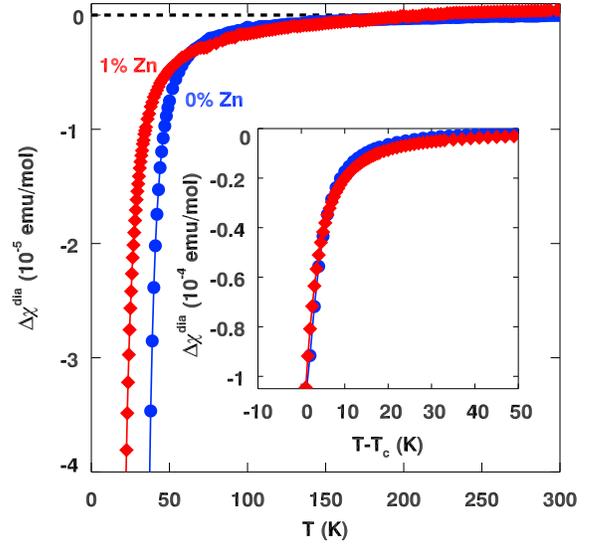}
\caption{(Color online) Fluctuation diamagnetism obtained using Eq.~ (\ref{eq:dchi}), based on bulk susceptibility measurements performed with a field of 100 Oe. Inset shows same data shifted by $T_c=32$~K for Zn0 and 17~K for Zn1. The unit emu/mol is equivalent to 4$\pi$cm$^3$/mol.}
\label{fig:fluct}
\end{figure}

We have shown previously\cite{huck08} that diamagnetic fluctuations appear well above $T_c$ in $\chi_{c}^s$ but not in $\chi_{ab}^s$; these are the same two-dimensional diamagnetic fluctuations that have been studied by Li, Ong, and coworkers with torque magnetometry.\cite{li10}  These fluctuations appear on top of the paramagnetic response, which scales as $g_i^2$.  Thus, to isolate the two-dimensional superconducting fluctuations, we plot in Fig.~\ref{fig:fluct} the quantity
\begin{equation}
  \Delta\chi^{\rm dia} = \left(g_{\rm ave}\over g_c\right)^2 \chi_c^{s}
     - \left(g_{\rm ave}\over g_{ab}\right)^2  \chi_{ab}^{s},
     \label{eq:dchi}
\end{equation}
where $g_{\rm ave}^2= (2g_{ab}^2+g_c^2)/3$.   As one can see, a weak diamagnetic response is apparent below 150~K, and it grows rapidly on approaching $T_c$.  The fluctuations are not significantly weakened by the Zn doping, but the divergent growth is shifted to a lower temperature.

\begin{figure}[b]
\includegraphics[width=0.8\linewidth]{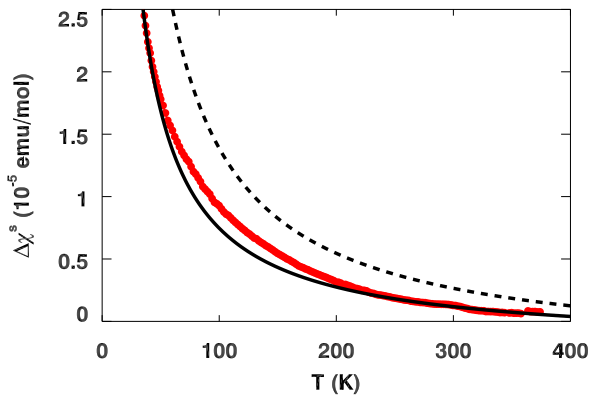}
\medskip
\caption{(Color online) Plot of $\Delta\chi^s$, as given by (\ref{eq:curie}), denoted by (red) circles.  Solid line is a calculated curve corresponding to a constant plus a Curie term (described in text), with the constant equal to $-2\times 10^{-6}$ emu/mol. The solid and dashed lines correspond to $S_{\rm eff}(S_{\rm eff}+1)=0.14$ and 0.25 respectively.}
\label{fig:curie}
\end{figure}

Another factor to examine is the paramagnetic susceptibility due to the Zn impurities.  To avoid the diamagnetic fluctuations, we focus on $\chi_{ab}^s$.  In Fig.~\ref{fig:curie} we plot
\begin{equation}
  \Delta\chi^s = \chi_{ab}^s({\rm Zn}1) -  \chi_{ab}^s({\rm Zn}0).
  \label{eq:curie}
\end{equation}
We have modeled this contribution with a small constant plus a Curie term, $Ng_{ab}^2\mu_{\rm B}^2S_{\rm eff}(S_{\rm eff}+1)/3k_{\rm B}T$, where $N$ is Avogadro's number and $S_{\rm eff}$ is the effective spin per Zn impurity. The curve shown in Fig.~\ref{fig:curie} corresponds to $S_{\rm eff}(S_{\rm eff}+1)=0.14$; the uncertainty in this quantity $\sim7$\%.

There have been a number of theoretical analyses of the effective spin per magnetic vacancy (equivalent to a Zn impurity) in various quantum spin systems.  For a system with an energy gap for singlet-to-triplet excitations, $S_{\rm eff}=\frac12$ and $S_{\rm eff}(S_{\rm eff}+1)=3/4$.\cite{sand97,sach99,vojt00}  For a 2D Heisenberg antiferromagnet in the dilute vacancy limit, $S_{\rm eff}(S_{\rm eff}+1) \approx 1/4$, shown as the dashed line in Fig.~\ref{fig:curie}, which is considerably smaller.\cite{vojt00,hogl03,huck02}  Our result of 0.14 is smaller yet.  Of course, the 1\%~ Zn concentration corresponds to an average Zn separation of $\sim10a \approx 40$~\AA, which is smaller than the spin correlation length at low temperature.  When two Zn impurities sit on opposite sublattices within the same AF domain, their contributions to the spin susceptibility are negligible because the effective local impurity moments are antiparallel;  hence, impurity correlations reduce the effective spin per impurity.  Comparison with the models suggests that the the magnetic correlations above 50~K are not spin gapped, but instead involve striped antiferromagnetic correlations with a significant correlation length.

In this section, we have analyzed the susceptibility data for both Zn0 and Zn1. We have shown that i) there is a kink in the susceptibility curve at 27~K for Zn0, which corresponds to the completion of the LTO-LTLO transition; ii) the superconducting transition is reduced from 32~K to 17~K by the 1\% Zn doping; iii) 2D superconducting fluctuations exist in the normal state in both samples; iv) the effective spin per Zn impurity is considerably smaller than 1/2.

\subsection{Thermal conductivity and thermoelectric power}

\begin{figure}[b]
\includegraphics[width=0.8\linewidth]{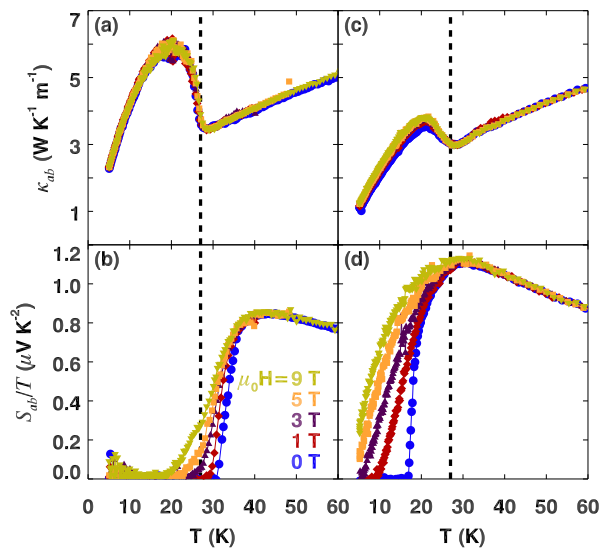}
\caption{(Color online)  Thermal conductivity measured parallel
to the planes, $\kappa_{ab}$, with field perpendicular to the planes for (a) Zn0, (c) Zn1.  Thermoelectric power, $S_{ab}$ divided by temperature $T$ for (b) Zn0, (d) Zn1.  Vertical dashed lines denote the characteristic temperature 27~K, as explained in the main text.}
\label{fig:thermal}
\end{figure}

Figure~\ref{fig:thermal}(a) and (c) show results for the $a$-$b$-plane thermal conductivity ($\kappa_{ab}$) for the Zn0 and Zn1 samples, respectively.  In both cases, $\kappa_{ab}$ is enhanced on cooling through 27~K, corresponding to the completion of the transition to the LTLO phase.   Together with the absence of any significant change with $H_\bot$, $\kappa_{ab}$ appears to be dominated by phonons.  In studies of related materials,\cite{babe98,sun03} it has been proposed that ordering of charge stripes could decrease the scattering rate of phonons, thus increasing the thermal conductivity.  In the present case, such an explanation seems unlikely as the stripe ordering is weak, while the change in $\kappa_{ab}$ is substantial, especially in comparison with comparably doped \lsco.\cite{naka91}  Instead, we point out that the change in $\kappa_{ab}$ correlates with the fall off in the critical lattice fluctuations indicated in Fig.~\ref{fig:td2}(c).  The reduced $\kappa_{ab}$ in the Zn1 sample is consistent with increased phonon scattering due to the Zn dopants, as observed\cite{sun03b} in Zn-doped La$_2$CuO$_4$.

The ratio of the in-plane thermopower to the temperature, $S_{ab}/T$, is shown in Fig.~\ref{fig:thermal} (b) and (d) for the Zn0 and Zn1 samples, respectively.  The magnitude is similar for both samples in the normal state.  For Zn0, $S_{ab}/T$ turns over at $\sim40$~K and drops to zero approximately at $T_c$.  Reducing $T_c$ by applying $H_\bot$, $S_{ab}/T$ fans out to lower temperature; when it is finite at 27~K, as for a field of 9~T, the slope changes at lower temperature.  For Zn1, the turnover in $S_{ab}/T$ shifts to lower temperature, along with $T_c$.  In neither sample do we observe the change in sign of $S_{ab}$ that was found\cite{li07} in LBCO $x=1/8$ below 45~K.

\section{Discussion}

\subsection{Interplay between structure, superconductivity, and stripe order}

In LBCO $x=0.095$, the structural and superconducting transitions overlap and interact in interesting ways.  In zero field, the rapid downturn in $\rho_{ab}$ below 37~K is interrupted at 35~K, as shown in Fig.~\ref{fig:spike}(a), corresponding to the point where the LTO to LTLO transition begins in earnest, as indicated in Fig.~\ref{fig:td2}(a).  While slowed, the decrease of $\rho_{ab}$ picks up again, dropping sharply to zero at $T_c=32$~K, where the bulk diamagnetic response sets in.  Note that $\rho_{c}$, which shows insulating character above 40~K, turns over at $\sim37$~K and reaches zero at $T_c$.

For the sample studied in Ref.~\onlinecite{huck11}, the development of the diamagnetism [the drop in $\chi$ shown in Fig.~\ref{fig:spike}(b)] below $T_c$ is interrupted at 30~K, with a sharp minimum in the diamagnetism at 27~K, followed by recovery.  That particular crystal showed an LTO to LTLO transition that took place over a narrower temperature range, between 30~K and 27~K [see Fig.~13(a) of Ref.~\onlinecite{huck11}].  Related, but less dramatic, behavior is observed in the Zn0 sample studied here.  The structural transition begins at a higher temperature, but ends at the same point.  The features at 27~K in Fig~\ref{fig:spike} correspond with the thermodynamic completion of the transition as indicated by the sharp kink in $\kappa_{ab}$ shown in Fig.~\ref{fig:thermal}(a).  While comparing transition temperatures, it is worthwhile to note that our $x=0.095$ sample is different from that of the same nominal concentration studied by Dunsiger {\it et al.}\cite{duns08} and the $x=0.10$ measured by Adachi {\it et al.},\cite{adac05} as those crystals had structural transition temperatures of 45~K and 40~K, respectively.  The comparison indicates that our samples have a lower relative Ba concentration.\cite{huck11}

A feature at 27~K also appears in $\rho_c$ when we apply $H_\bot$.  For example, we see in Fig.~\ref{fig:spike}(a) that, for $\mu_0 H_\bot=3$~T, the initial decrease in $\rho_c$ below 37~K is interrupted at 30~K, with $\rho_c$ then rising to a peak at 27~K before resuming its decrease.  This decrease is remarkably gradual, considering that $\rho_{ab}$ is effectively zero below 30~K in the same field.  As discussed in paper I, such behavior, with finite resistivity in only one direction, violates conventional theoretical expectations.

The likely explanation for the anomalous changes in the development of the superconducting correlations is the sensitivity of the Josephson coupling between CuO$_2$ planes to the structural transition.  Disruption of the interlayer Josephson coupling by the LTO to LTLO transition was first pointed out in a study of La$_{1.85-y}$Nd$_y$Sr$_{0.15}$CuO$_4$ by Tajima {\it et al.}\cite{taji01}  A strong suppression of Josephson coupling has also been identified in the LTLO phase of LBCO $x=1/8$.\cite{li07,scha10b}

In the present situation, the effect is more subtle. Strong superconducting correlations begin to develop when the crystal is largely LTO.  As the volume fraction of the LTLO phase increases, the average Josephson coupling decreases.  At 27~K, where the transition to the LTLO phase is effectively complete, the Josephson coupling presumably reaches its minimum value, which it retains on further cooling.

The bulk diamagnetism is disrupted by the structural transition because the superconducting order parameter is sensitive to the 3D coupling.  When the Josephson coupling rapidly decreases on cooling to 27~K, the degree of superconducting order is reduced, and one must cool to a significantly lower temperature to achieve full order. Further evidence for the temperature-dependent evolution of the Josephson coupling comes from optical studies of the c-axis reflectivity.\cite{home12}

From discussions above, it is now clear that the LTLO transition results in a partial frustration of the $c$-axis Josephson coupling, which causes disruption on the development of the bulk superconductivity. Upon completion of the transition, weak stripe order develops. Under an external magnetic field, the stripe order is enhanced. Interestingly, while the enhanced stripe order correlates with weakened interlayer coupling, the in-plane superconductivity remains robust, as discussed in detail in paper I.

For the state of uniaxial resistivity in finite $H_\bot$, we have presented evidence in paper I that $\rho_c$ can be finite even when the interlayer
Josephson coupling is finite.  Here we emphasize that the 27-K peak in $\rho_c(T)$ for finite $H_{\bot}$ can be understood in terms of the sensitivity of that Josephson coupling to the crystal symmetry.  The situation is related to, but quantitatively different from, the case of LBCO $x=1/8$, where the stripe order is strong and the frustration of the Josephson coupling is quite substantial.\cite{li07,tran08}  The pair-density-wave (PDW) state, with sinusoidal modulation of the superconducting pair wave function associated with stripe order, has been proposed\cite{hime02,berg07} to explain the frustrated Josephson coupling.  (Recent calculations\cite{corb11,lode11b} provide evidence for the energetic favorability of such a state.)  For $x=0.095$, the development of a spatially modulated pair density\cite{baru08} would help to explain the structure-sensitivity of the Josephson coupling; the frustration may become substantial when the field is large.

An interesting observation is that the effective Josephson couplings, as measured by the Josephson plasma resonance (JPR), for LBCO and LSCO with $x\sim0.1$ are almost identical at low T,\cite{scha10b} despite the fact that LSCO remains in the LTO structure.  Field-induced spin-stripe order\cite{lake02} has been invoked to explain the rapid reduction of the JPR of underdoped LSCO in $H_\bot$.\cite{scha10}    Even in zero-field, there has been evidence for close proximity to stripe ordering in LTO-phase LSCO.\cite{hunt99,yama98a} Raman scattering studies also provide evidence for charge-ordering fluctuations in underdoped LSCO.\cite{musc10}  Hence, our LBCO $x=0.095$ sample seems to be tuned to a regime in which the Josephson coupling is especially sensitive to the change in lattice symmetry and the degree of stripe correlations.

Given our analysis of the LBCO $x=0.095$ sample, it is worthwhile to reconsider the results obtained previously for $x=1/8$.  There an inflection point in $\rho_{c}$ was observed near the large drop in $\rho_{ab}$ at $\sim40$~K, with $\rho_c$ starting to decrease below 35~K.\cite{li07,tran08}  These changes indicate that there must be some weak Josephson coupling between the layers, despite the lack of superconducting order.   When $\rho_{ab}\rightarrow0$ at 16~K, $\rho_c$ has a substantial magnitude but is significantly reduced compared to the normal state.  Thus, the phase obtained in low magnetic field for $x=1/8$ appears to be similar to what we find at high fields for $x=0.095$ (see paper I) with effectively zero resistivity parallel to the planes but dissipation in the perpendicular direction.

\subsection{Impact of Zn substitution}

By substituting 1\% Zn for Cu in LBCO $x=0.095$, the bulk \tc is reduced to 17~K, and the screening volume fraction is also slightly reduced, as shown in Fig.~\ref{fig:sc}.  At the same time, the spin-stripe order is enhanced, as shown in Fig.~\ref{fig:spin}(b), with the magnetic peak intensity rising below $\sim27$~K in a fashion similar to a second-order phase transition.   This is consistent with the Curie-Weiss analysis of the susceptibility data, which suggests the presence of substantial spin correlations in the paramagnetic phase.  At low temperature, the magnitude of the spin-stripe order is certainly far from saturated, as $H_\bot$ enhances the intensity.  We note that an initial attempt to detect the charge stripe order in the Zn1 sample by x-ray diffraction was unsuccessful.\cite{huck11b}

It is interesting to compare the impact of Zn dopants with that of vortices.  On the basis of muon-spin-rotation measurements, Nachumi {\it et al.}\cite{nach96} argued that each Zn dopant depletes the superfluid density in an area comparable to that of a magnetic vortex core.  These are also the regions that presumably pin stripes, with finite correlations between these regions leading to enhanced order.\cite{lake01,lake02,deml01,chen02,kive02}
Some aspects of these results are captured by recent calculations for non-magnetic impurities\cite{ande07b,ande10} and magnetic vortices\cite{schm10} in a disordered $d$-wave superconductor.

\subsection{Thermopower}

Thermopower is a useful experimental property, as it is straightforward to measure; however, it can be challenging to interpret.  The thermopower in the normal state of underdoped cuprates is positive, decreasing in magnitude as the hole concentration grows.\cite{ober92}  This trend is disrupted in stripe-ordered materials.  Measurements on La$_{2-x-y}$Nd$_y$(Ba,Sr)$_x$CuO$_4$ (Ref.~\onlinecite{naka92,yama94b,bao96b,huck98,xie11}), La$_{1.8-x}$Eu$_{0.2}$Sr$_x$CuO$_4$ (Ref.~\onlinecite{huck98,chan10}), and \lbco\ (Ref.~\onlinecite{adac01,li07}) show that $S/T$ decreases rapidly on cooling below the LTLO transition, eventually going negative.  A sign change in $S/T$  (and in the Hall resistance\cite{lebo11}) has also been observed in YBCO crystals with hole concentrations near 1/8.\cite{lali11,chan10}  The negative sign of $S$, together with observations of quantum oscillations, has been interpreted as evidence for electron pockets that occur due to Fermi surface reconstruction associated with stripe order.\cite{lali11,chan10}  Fairly direct evidence for charge-stripe order induced in underdoped \ybco\ by high magnetic field has recently been provided by a nuclear magnetic resonance study.\cite{wu11}

In the present case of LBCO with $x=0.095$, we have seen (in paper I) that a $c$-axis magnetic field enhances both charge and spin stripe order.  While we do see a downturn in $S_{ab}/T$ starting about 10~K above $T_c$, there are no indications of a sign change in $S_{ab}/T$.  The downturn is likely due to superconducting fluctuations, as discussed in the next section. There appears to be robust in-plane superconductivity in this sample, and the thermopower of the superconducting state is zero. [The apparent rise in $S_{ab}/T$ below 10~K in Fig.~\ref{fig:thermal}(b) is simply an enhancement of noise in $S_{ab}$ when divided by small $T$.]  One needs to suppress the superconductivity in order to see whether the recovered normal state has negative thermopower.  Doping with Zn is one way to depress the superconductivity, and in our Zn1 sample we have also seen an enhancement of spin stripe order in zero field.  Below 50~K, the $S_{ab}/T$ for Zn1 appears to be an extrapolation of the high-temperature behavior of the Zn0 sample, again with a downturn starting about 10~K above the reduced $T_c$.  Electronic scattering by the Zn would have some impact on transport, but it is not clear whether that would impact the sign of $S_{ab}$.

The electron-pocket interpretation of negative thermopower requires that there be a finite density of states at the Fermi level in the antinodal region of reciprocal space, the same region where angle-resolved photoemission spectroscopy (ARPES) typically indicates a substantial energy gap, even in the normal state.\cite{dama03}  Unfortunately, it is not possible to use ARPES to test for electron pockets in the high magnetic fields required to achieve stripe order in YBCO\cite{wu11}; however, it has been possible to do ARPES measurements on stripe-ordered LBCO\cite{vall06,he09} in the same temperature regime where negative thermopower is observed.  As discussed elsewhere,\cite{tran10b} those ARPES results are inconsistent with antinodal electron pockets.

Given the situation, it seems worthwhile to reconsider the interpretation of the thermopower.  In general, the thermopower measures the entropy per charge carrier.  In the low-temperature limit, $-S/T$ is proportional to the derivative of the electronic density of states with respect to energy, evaluated at the chemical potential.\cite{mats11b}  In layered cuprates, the dominant contribution to the electronic density of states comes from the antinodal region of reciprocal space because of relatively flat band dispersion in that region.\cite{dama03}  With increasing doping, the associated van Hove ``singularity"
 eventually shifts from below to above the Fermi level, $E_F$, changing the sign of the energy derivative of the density of states.   To test the connection with $S$,  Kondo {\it et al.}\cite{kond05} performed both ARPES and thermopower measurements on a series of (Bi,Pb)$_2$(Sr,La)$_2$CuO$_{6+\delta}$ crystals ranging from optimally to highly over-doped.  They found that $S/T$ changes sign from positive to negative as the measured van Hove singularity moves from below to above $E_F$.

Kondo {\it et al.}\cite{kond05} found that a quantitative evaluation of the thermopower based on measured electronic dispersions broke down for a slightly underdoped sample.  Interpretations of negative thermopower for YBCO based on Fermi-surface reconstruction make use of Fermi-liquid assumptions,\cite{lali11,chan10} which are rather dubious for the underdoped regime.  For the sake of argument, we wish to suggest another possibility.  The key feature is still associated with the antinodal states that tend to dominate the thermopower.  The pair-density-wave (PDW) superconductor proposed to develop on top of charge and spin stripe order\cite{hime02,berg07} should have a particle-hole symmetric gap in the antinodal regions, together with gapless arcs.\cite{baru08}  The antinodal gap would cause the contribution of the antinodal states to the thermopower to go to zero, with the residual response coming from the nodal arcs.\footnote{A particle-hole symmetric pseudogap should result in a partial reduction of the thermopower.  In underdoped \bscco, there is evidence for such a pseudogap at 140~K from ARPES\cite{yang11} as well as considerable rounding of $S(T)$ well above $T_c$.\cite{fuji02e}}  Whether the arc states could provide a negative contribution to the thermopower requires an evaluation beyond our capabilities.  In any case, we suggest that the rapid drop of $S_{ab}/T$ seen coincident with charge stripe order in low-temperature-tetragonal (LTT)-structured cuprates with $x\approx 1/8$ could be due to a subtle change in the antinodal states, with the higher-temperature antinodal pseudogap switching to a gap that is symmetric in energy about $E_F$.  When a $d$-wave-like gap opens up on the arcs below 40~K,\cite{he09} the thermopower drops to zero.\cite{tran08}  In the case of $x=0.095$, bulk superconductivity is already present when stripe order begins to develop, and the evolution of the superconducting order determines the low-temperature thermopower.

\subsection{Superconducting fluctuations above \tc}

The nature of superconducting fluctuations and their relevance to determining \tc\ in underdoped cuprates continue to be contentious issues.  In BCS theory, it is assumed that superconducting order is limited only by the development of pairing correlations, with phase coherence following as soon as the pairing amplitude develops.\cite{bard57}  In contrast, Emery and Kivelson\cite{emer95a} proposed that \tc\ of underdoped cuprates is limited by fluctuations of the phase of the superconducting order parameter.  Superconducting fluctuations close to \tc\ have been considered for some time,\cite{ulla91} but the discussion shifted when Ong and collaborators\cite{xu00,wang01,wang02,wang06} identified an anomalously large and positive contribution to the Nernst effect that appears far above \tc, and attributed it to fluctuations of superconducting vortices.  More direct evidence for superconducting fluctuations has come from torque magnetometry studies,\cite{li05,li07b,li10} which indicate an anisotropic diamagnetic response extending well above \tc.  This work has stimulated analyses of vortex fluctuations in the ``normal'' state.\cite{lee06,ande08}

Measurements of dynamical conductivity have led to slightly different conclusions.  While such studies do support the idea that the loss of superconducting order is due to phase fluctuations within the CuO$_2$ planes,\cite{cors99,ohas09} they also suggest that superconducting contributions to the conductivity disappear within $\sim10$~K above \tc.\cite{bilb11a}  Bilbro and coworkers\cite{bilb11b} have compared the relative strengths of superconducting fluctuations in magnetic susceptibility and electrical conductance with theoretical predictions for a 2D system with superconducting order limited only by vortex fluctuations.  They find that the ratio is off by almost an order of magnitude just above \tc.  In a recent study of diamagnetism above $T_c$ in LSCO $x=0.1$, Mosqueira {\it et al.}\cite{mosq11} have argued that their observations can be explained by chemical disorder plus fluctuations in the pairing amplitude.  At minimum, these results cast doubt on a picture of the normal state involving decoupled layers with superconductivity limited only by phase fluctuations.

In light of this previous work, we consider the present results.  We have seen in Fig.~\ref{fig:fluct} that fluctuation diamagnetism becomes detectable at $\sim100$~K above $T_c$.  It is not weakened by 1\%\ Zn doping, although the divergence of the diamagnetic response shifts with the bulk \tc, as indicated in the inset of Fig.~\ref{fig:fluct}.  The lack of sensitivity to weak impurity substitution for $T>T_c +30$~K suggests that the associated correlation length for the diamagnetic response is smaller than the average impurity spacing, which is approximately 40~\AA.  As the correlation length for the phase of the superconducting order parameter begins to grow more rapidly below 60~K, the impact of the Zn impurities becomes apparent.

We can detect small levels of diamagnetism due to its anisotropic dependence on the direction of the magnetic field.  This is the same anisotropy that is exploited in the torque magnetometry studies.\cite{li07b}  The situation is different when we look at transport properties.

\begin{figure}[t]
\includegraphics[width=0.8\linewidth]{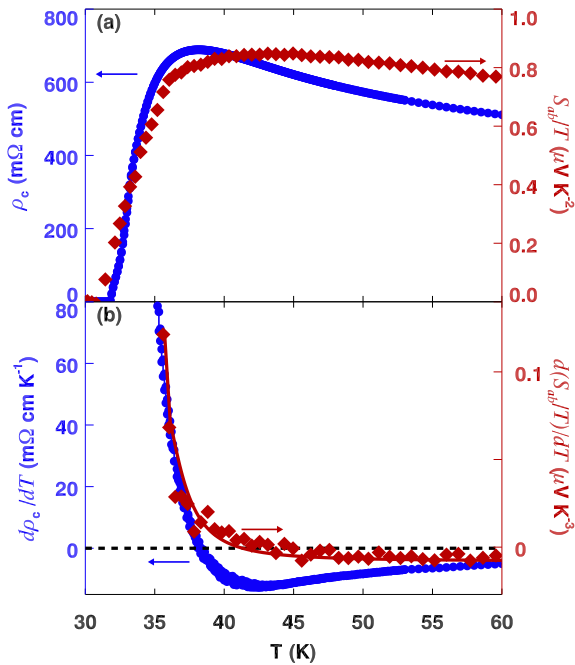}
\caption{(Color online)  (a) Comparison of $\rho_c(T)$ (circles, data from Ref.~\onlinecite{wen11}) and $S_{ab}/T$ [diamonds, from Fig.~\ref{fig:thermal}(b)] for Zn0 in zero magnetic field.  (b) Temperature derivatives of  $\rho_c(T)$ (circles) and of $S_{ab}/T$ (diamonds). }
\label{fig:deriv}
\end{figure}

We have already suggested that the downturn in $S_{ab}/T$ is due to superconducting fluctuations.  This is reinforced by the sensitivity to $H_\bot$ in this region, as indicated in Fig.~\ref{fig:thermal}(b).  As another measure, we plot the derivative of $S_{ab}/T$ in Fig.~\ref{fig:deriv}(b).  The change in sign of the derivative at $\sim42$~K provides a measure of the onset of significant superconducting fluctuations.  We compare with the derivative of $\rho_c$, which has a minimum at the same temperature, corresponding to an inflection point for $\rho_c(T)$.  That is also the approximate point below which $H_\bot$ causes $\rho_c$ to increase, as shown in paper I.    While the magnitude of $\rho_c$ is very large at these temperatures, the only plausible reason for it to change its curvature and become sensitive to field is because of superconducting fluctuations involving pair tunneling between the CuO$_2$ planes.\cite{moro00}  Thus, we appear to have indications of 3D superconducting fluctuations at $T_c+10$~K.

We have argued earlier that even for $x=1/8$ the effects of Josephson coupling appear before long-range superconducting phase coherence is established within the planes.  It appears to be difficult to realize decoupled superconducting layers with the degree of correlations limited only by phase fluctuations.  The arguments of Bilbro {\it et al.}\cite{bilb11b} indicate that this is problematic even very close to $T_c$.  One possibility is that disorder plays an essential role for superconducting correlations above $T_c$.  Scanning tunneling spectroscopy studies\cite{gome07,pasu08} of \bscco\ have indicated a granular character to superconducting correlations above $T_c$, and a scanning magnetometry study\cite{iguc01} of an LSCO film found inhomogeneous magnetic domains that survived to $T_c+60$~K.  The electronic inhomogeneity may be a necessary ingredient to explain the large magnitude of $\rho_{ab}$ in the normal state.  Indeed, Schneider and Weyeneth\cite{schn11} have recently presented an analysis of the diamagnetism and Nernst signal based on a granular picture.  Similarly, there has been an estimate of the diamagnetism from fluctuating stripes.\cite{mart10} It would be interesting to see an evaluation of resistivity together with diamagnetism for the same model.

The experimental regime of most extreme contrast is in LSCO with $x\lesssim0.055$.  Li {\it et al.}\cite{li07b} found evidence for diamagnetism here, despite the fact that the samples are insulating at low temperature.\cite{ando01}  Intriguingly, Bollinger {\it et al.}\cite{boll11} recently presented evidence for a quantum critical point at $x\approx0.055$, with a Bose insulator phase at lower doping.  Again, there are also stripes, though of a different breed.\cite{birg06,gran04}  Might all of these phenomena be connected?

\section{Conclusions}

We have characterized the structural transition from the LTO to LTLO phase and its impact on the superconducting transition in LBCO with $x=0.095$.  On cooling, the divergence of the superconducting correlation length is interrupted by the onset of the structural transition, as indicated by a shoulder in $\rho_{ab}(T)$.  The apparently first order transition takes approximately 8~K to complete, with its effective termination at 27~K indicated by a sharp change in the thermal conductivity. Weak stripe order also appears in this region.  We have argued that the structural transition continuously weakens the interlayer Josephson coupling, with clear impacts reflected in the bulk susceptibility and in $\rho_c(T)$ measured in $H_\bot$.  One possible reason for the strong sensitivity of the bulk superconductivity to the structure would be spatial modulation of the superconducting order parameter, especially a modulation in phase with the charge stripe order.

Substituting 1\%\ Zn for Cu has minimal effects on the structural transition, but significantly reduces the superconducting $T_c$. Zn doping and $H_\bot$ each enhance the spin stripe order.  Analysis of the anisotropic magnetic susceptibility of samples with and without Zn suggests that the Zn impurities cause the ordering of Cu spins that have a substantial correlation length in the absence of the impurities, consistent with the existence of dynamical stripe correlations.

We have compared thermopower measurements on the present samples with the anomalous behavior found in LBCO $x=1/8$.  We have argued that the transition to stripe order and negative thermopower in the latter case might be explained by the development of a particle-hole symmetric gap for antinodal states, as predicted for a PDW superconductor.  We have also examined evidence for superconducting fluctuations above $T_c$.  Anisotropic diamagnetism extends 100~K above $T_c$, whereas the effects in the thermopower and $\rho_c$ are only evident to $\sim10$~K above $T_c$.  It appears that disorder may have a significant role in modifying the signatures of superconducting fluctuations in transport properties relative to magnetic susceptibility.

\begin{acknowledgements}
We gratefully acknowledge helpful comments from R. Konik.
The work at Brookhaven was supported by the Office of Basic Energy Sciences, Division of Materials Science and Engineering, U.S. Department of Energy (DOE), under Contract No.\ DE-AC02-98CH10886. JSW, ZJX, and (in part) JMT were supported by the Center for Emergent Superconductivity, an Energy Frontier Research Center funded by the U.S. DOE, Office of Basic Energy Sciences.  Research at Oak Ridge National Laboratory High Flux Isotope Reactor was sponsored by the Scientific User Facilities Division, Office of Basic Energy Sciences, U.S. Department of Energy. SPINS at NCNR is supported by the National Science Foundation under Agreement No.\ DMR-0454672.
\end{acknowledgements}


%

\end{document}